\documentclass[11pt]{article}
\usepackage{amsmath, amssymb, physics}
\usepackage{geometry}
\usepackage{authblk}
\usepackage{graphicx}
\usepackage{bm}
\usepackage{url}
\usepackage{color}
\usepackage{float}
\usepackage{graphics}
\usepackage{tablefootnote}
\usepackage{mathtools}
\usepackage{cite}
\title{Thermodynamics of Innovation: A Statistical Mechanics Framework of Social Adoption}
\author{Guilherme S. Y. Giardini$^{a}$, Carlo R. daCunha$^{a}$ 
\\
$^a$ Sanghi College of Engineering (SCE) and School of Informatics, Computing, and Cyber Systems (SICCS), Northern Arizona University, Flagstaff, AZ, 86011, US.
}
\date{}

\begin{document}
\maketitle

\begin{abstract}
    We develop a thermodynamic framework for modeling innovation adoption and abandonment dynamics using statistical mechanics. Starting from a mathematical model for an adoption distribution that fits empirically obtained date, we construct a canonical ensemble whose equilibrium distribution yields Gompertz-like and Maxwell–Boltzmann–like shapes. By reverse-engineering the associated energy landscape, we define an effective potential and derive a dynamical Lagrangian formulation. The resulting field theory captures key features of emergent behaviors in socio-technical systems, from early suppression to peak dynamics and late decline. We interpret effective temperature, entropy, and equilibrium points, and show how these systems exhibit hybrid thermodynamic-statistical signatures.
\end{abstract}

\section{Introduction}
Adoption and abandonment processes, whether in technology diffusion, biological populations, or socio-economic trends—often exhibit asymmetric, non-Gaussian, and skewed temporal distributions. Traditional models like the one proposed by Bass \cite{Bass2001,Lee2014,Massiani2015,Tchouya2023}, popularly known as the S-curve of adoption, shows a initial slow uptake in technology adoption throughout a population until adoption gains momentum as more people adopt the idea and starts to spread that technology by imitation, up to the point where it reaches saturation. This type of modeling has very good prediction capabilities, and is at many times used in product cycle, just in time production, and many other industrial, and commercial applications \cite{Guo2014,Han2025}. While the s-shape models capture many different adoption dynamics, they only model the uptake dynamics of innovation, as competition arises, and groups of people get saturated by an idea, or competition of technologies happens, an idea or technology can be forgotten. Thus for a more precise type of modeling, a more comprehensive model that encompasses rise-and-fall dynamics would be ideal \cite{Lee2022}. 

Another important characteristic of innovation models and opinion models, is that many of the models are inspired from magnetic system theory like spin glass systems and Ising models, where magnetization represents an individual state, and this state, in a sociological analogy may represent opinions, technology adoption, or ideas \cite{Vandermaas2020,Baldassarri2023,Korbel2023,Macy2024}. Also, it is a known fact that thermodynamic theory can be applied to magnetic systems, and provide important insights on how they behave under temperature variations, energy variations, how susceptible they are to external influences and so on. Since thermodynamics is such a powerful tool, and is compatible with magnetic systems, it is a given that it should also be applied to innovation and other sociological systems. While some studies have tried to employ pure thermodynamics to social systems, they cannot quantify more complex dynamics, since thermodynamics is a theory that works by assuming averages \cite{Marsland2017, Vargo2020}. 

Here, we propose a new formulation by constructing a thermodynamic system whose canonical distribution exactly reproduces the innovation dynamics proposed in \cite{Giardini2024}, where adoption and abandonment dynamics are inspired by chain reactions of atomic decays. Such approach allows studying innovation from a dynamical approach with differential equations, but also allows an averaged theory of thermodynamics, that leads to a vision of innovation through macroscopic lenses of temperature, entropy, and innovation potentials. 

\section{Empirical Distribution and Model Motivation}

We represent innovation dynamics as a Markovian stochastic birth–death process, conceptually related to radioactive decay and spin-flip mechanisms in magnetic systems, and extended to incorporate adoption–abandonment trade-offs \cite{Giardini2024}. Let $X_{n}(t)$ denote the probability (or fraction) of $n$ agents being in the adoption state for a given idea or technology. Agents transition between adoption and abandonment with time-inhomogeneous rates $\lambda_{n}(t)$ and $\mu_{n}(t)$, respectively, corresponding to ``birth'' and ``death'' processes in the adoption space.

A mean-field type simplification is introduced, assuming that these rates depend solely on the total number of adopters, so that $\lambda_{n}(t) = n \lambda_{0}$ and $\mu_{n}(t) = n \lambda_{0}(t)$. To reflect persistence: agents' resistance to changing previously held positions, we include a survival term $\left[ 1 - (\lambda_{n} + \mu_{n})\Delta t \right] X_{n}(t)$. Adoption over the interval $\Delta t$ is described by the gain term $\lambda_{n-1} \Delta t \, X_{n-1}(t)$, while abandonment is modeled through the loss term $\mu_{n+1} \Delta t \, X_{n+1}(t)$.

To find the adoption rate $\lambda_0(t)$ we considered a set of independent opportunities $i$ that independently can lead to adoption events. Each of the opportunities have an intrinsic realization rate $\Lambda_i>0$, and once realized it is depleted. 

The overall birth process, therefore, must be the superposition of these independent events. We assume that opportunity realizations follow a Poisson process. Therefore, the probability that an opportunity $i$ remains unrealized by time $t$ is $e^{-\Lambda_it}$, so the instantaneous rate at which it contributes to a birth at time $t$ is $\Lambda_ie^{-\Lambda_it}$.

The total birth rate at time $t$ is then:

\begin{equation}
	\lambda_0(t)=\sum_i\lambda_ie^{-\Lambda_it}\rightarrow\int_0^\infty\Lambda f(\Lambda)e^{-\Lambda t}d\Lambda,
	\label{eq:um}
\end{equation}
where we assumed a continuum limit, with $f(\Lambda)$ being the density of of opportunities. To find this density, we assume that each opportunity have an energy cost $E>0$ to realization, leading to an Arrhenius-like rate $\Lambda=\nu e^{-\beta E}$, so that $d\Lambda/\Lambda=-\beta dE$. If we assume that the barriers are broadly distributed without a preferred scale as in a disordered system, we can assume that the density is $\rho(E)=1/E_{max}$. Consequently, the number of oppotunities in $E$-space is $\rho(E)dE$, while in $\lambda$-space it is $f(\lambda)d\lambda$. Therefore:

\begin{equation}
	\frac{1}{E_{max}}dE=f(\Lambda)d\Lambda\rightarrow f(\Lambda)\propto\frac{1}{\Lambda}.
\end{equation}
Consequently, Eq. \ref{eq:um} leads to:

\[
    \lambda_{0}(t) = \frac{\alpha}{t},
\]
which plays the role of a \emph{birth-ratio} over time, with $\alpha$ serving as a characteristic adoption time constant. The hyperbolic decay encapsulates the intuition that opportunities for ``first adoptions'' diminish as the system ages, reflecting a kind of critical slowing down often observed in natural and social processes that display threshold-like behavior \cite{Laibson1998,Rogers2003}. In contrast, abandonment is governed by a mean-reverting process that captures agents exiting a technology at a rate that fluctuates around a mean due to systemic pressures. Specifically, we model:

\[
    \mu_{0}(t) = \beta (1 - \exp(-t/\sigma)) \, .
\]
after the Ornstein-Uhlenbeck (OU) form \cite{Uhlenbeck1930, daCunha2022}. This choice reflects the \emph{reversibility of commitment}: while agents may initially resist abandoning, the probability grows and saturates toward a baseline $\beta$, capturing both persistence at early stages and eventual stabilization of dropout rates. Together, the hyperbolic adoption kernel and the O-U abandonment kernel form a complementary pair: one encoding the fading novelty of innovation; the other encoding the gradual settling into stable patterns of desertion.

This formulation mirrors diffusion processes, but here the ``diffusion'' occurs in an abstract adoption space: agents stochastically move between adopter and non-adopter states rather than through physical positions. Constructing the dynamical equation for $X_{n}(t)$ with these persistence, adoption, and abandonment components, and summing over all $n$ to define the macroscopic quantity:
\[
M(t) = \sum_{n=1}^{\infty} X_{n}(t),
\]
analogous to magnetization in statistical physics, yields a Gompertz-type functional form for the adoption–abandonment curve \cite{Gompertz1997}:
\begin{equation}
M(t) = M_{0} \, t^{\alpha} \, \exp\!\left[ -\beta \left( \sigma e^{-t/\sigma} + t \right) \right].
\end{equation}

This expression effectively captures empirical patterns across diverse domains, including citation trajectories, online platform adoption, and the rise-and-decline cycles of technologies such as 3D televisions and portable game consoles like the Playtation Portable (PSP). The recovered dynamics span both gradual, Socratic-style diffusion, and disruptive Schumpeterian innovation waves \cite{Lee2022}.

In earlier work, we interpreted the independent variable $t$ strictly as chronological time \cite{Giardini2024}. Here, we take a broader perspective: the same mathematical structure can describe adoption and abandonment dynamics as a function of a \emph{generalized state variable} $\omega$, which does not need to be time. In particular, we focus on the case where $\omega$ corresponds to an \emph{exposure frequency} (the cumulative number of relevant encounters, events, or interactions experienced by the population). In this framing, each increment in $\omega$ corresponds to an exposure event rather than a fixed time step. Adoption accelerates at intermediate $\omega$ due to increasing awareness, while high $\omega$ leads to saturation and abandonment as novelty decays. The mathematical form of the model remains identical, but its interpretation shifts from being time-driven to exposure-driven.

The generalized adoption–abandonment profile is thus written as:
\begin{equation}
M(\omega) = M_0 \cdot \omega^{\alpha} \cdot \exp\left( -\beta\left[ \sigma e^{-\omega/\sigma} + \omega \right] \right),
\label{eq:Innovation_Dynamics}
\end{equation}
where $\alpha$ is the adoption coefficient, $\beta$ the abandonment (or decay) coefficient, and $\sigma$ a characteristic suppression scale. While we emphasize the interpretation of $\omega$ as exposure frequency, the same functional form can describe other monotonically increasing state variables such as cumulative influence, resource depletion, or activation energy.

\subsection{Innovation Function as a Generalized Maxwell-Boltzmann Equation }

Equation \ref{eq:Innovation_Dynamics} has a Maclaurin expansion:
\begin{equation}
	M(\omega) \simeq M_0'\,\omega^{\alpha}\,\exp\!\left[-\frac{\beta}{2\sigma}\,\omega^2\right]\exp\!\left[ \frac{\beta}{6\sigma^2}\,\omega^3 + \dots\right].
	\label{eq:Taylor}
\end{equation}
This equation reduces to the Maxwell–Boltzmann distribution \cite{Maxwell1860,Boltzmann1872}, with Gaussian core and density-of-states factor $\omega^{\alpha}$ when $\alpha = 2$ and higher-order terms corresponding to anharmonic corrections\footnote{In analogy with classical mechanics, ``harmonic'' refers to a purely quadratic potential in the exponent, here producing the Gaussian core of the Maxwell–Boltzmann distribution. 
``Anharmonic'' corrections are higher-order terms ($\propto \omega^3, \omega^4, \dots$) in the exponent that deform the Gaussian shape. 
In this model they arise naturally from the series expansion of the original Gompertz term, and are suppressed by powers of $(\omega/\sigma)^3$ and higher.} are discarded. Higher-order terms can be discarded in the limit of large $\sigma$, such that the dimensionless ratio $\omega/\sigma$ remains small, regardless of whether $\omega$ represents time, exposure frequency, or another monotonically increasing progress coordinate.

Physically, the regime $\sigma\gg\omega$ corresponds to \emph{rapid decline in adoption with respect to the chosen state variable $\omega$}. For example, in the exposure-frequency interpretation, this is the regime where abandonment dominates after only a modest number of exposures.

\section{Canonical Formulation}

Since the adoption–abandonment profile \eqref{eq:Innovation_Dynamics} converges to a Maxwell–Boltzmann–type distribution for large $\sigma$ values, we can recast the model in the language of equilibrium statistical mechanics. This analogy allows us to introduce the concepts of an effective innovation energy with respect to the underlying variable $\omega$.  
We introduce a canonical ensemble over the state variable $\omega$, with probability density:
\begin{equation}
P(\omega) = \frac{1}{Z}\, g(\omega)\, e^{-\beta E(\omega)}\,,
\label{eq:canon_dist}
\end{equation}
where $Z$ is the partition function ensuring normalization. We set $\beta' = \beta\, m$, where $\beta$ is the inverse temperature and $m$ is a mass-like scaling parameter, directly analogous to the particle mass in the Maxwell–Boltzmann distribution. In the present context, $m$ globally scales the abandonment tendency with respect to $\omega$: larger $m$ produces narrower distributions peaked at smaller $\omega$, while smaller $m$ yields broader curves with peaks at larger $\omega$.

The degeneracy factor is taken as:
\begin{equation}
g(\omega) = \omega^{\alpha},
\end{equation}
representing the entropic multiplicity of microstates accessible at a given $\omega$. The exponent $\alpha$ is inherited directly from the underlying innovation dynamics and quantifies how the number of accessible configurations grows with the state variable.

The effective energy landscape, extracted directly from Eq. \ref{eq:Innovation_Dynamics} is defined as:
\begin{equation}
E(\omega) = m\,\omega + m\,\sigma\,e^{-\omega/\sigma},
\end{equation}
with two distinct contributions: a linear term, $m\,\omega$, representing a constant abandonment pressure that penalizes large $\omega$, and an exponential term, $m\,\sigma\,e^{-\omega/\sigma}$, encoding an early-stage suppression that diminishes with increasing $\omega$. The parameter $\sigma$ sets the crossover scale between this initial barrier (not a potential barrier, but a process of hitting critical mass for adoption \cite{Rogers2003}) and the asymptotic decay, while $m$ controls the global intensity of abandonment. The inverse temperature $\beta = \beta'/m$ weights states according to their energy in the usual Boltzmann factor.

In this mapping, $\alpha$ enters purely through the entropic degeneracy $g(\omega)$, $\beta$ through the thermal weighting, and $(m, \sigma)$ through the structure of the potential $E(\omega)$. The canonical partition function is therefore:
\begin{equation}
Z(\alpha,\beta,m,\sigma) =
\int_{0}^{\infty} \omega^{\alpha}
\exp\!\left[-\beta\, m \left(\omega + \sigma\, e^{-\omega/\sigma} \right) \right]\,d\omega,
\label{eq:partition_function}
\end{equation}
which defines a valid, normalizable probability distribution for $\alpha > -1$ and $\beta, m, \sigma > 0$. The formulation also holds in the limit $\sigma \to \infty$, which reduces the adoption curve to the Maxwell–Boltzmann.

An exact series representation follows from expanding the Gompertz term:
\begin{equation}
Z = \Gamma(\alpha+1)\sum_{n=0}^{\infty}
\frac{[-(\beta m \sigma)]^{n}}{n!}\,
\left(\beta m + \frac{n}{\sigma}\right)^{-(\alpha+1)},
\label{eq:Z_series}
\end{equation}
which converges absolutely for all admissible parameters.  
In the large $\sigma$ (Maxwell–Boltzmann) limit, Eq.~\eqref{eq:partition_function} reduces asymptotically to:
\begin{equation}
Z \;\approx\; \frac{1}{2}\,e^{-\beta m\sigma}\,
\Gamma\!\left(\frac{\alpha+1}{2}\right)
\left(\frac{2\sigma}{\beta m}\right)^{\frac{\alpha+1}{2}},
\quad \beta m \sigma \gg 1.
\label{eq:Z_MB_limit}
\end{equation}

\section{Effective Potential and Lagrangian Dynamics}

The canonical energy function \(E(\omega)\) defines an effective potential landscape for the generalized state variable \(\omega\) once thermal weighting and entropic degeneracy are incorporated.  
Including the Boltzmann factor and degeneracy term \(g(\omega) \propto \omega^{\alpha}\), the dimensionless effective potential reads:
\begin{equation}
    V(\omega) = \beta m\,\omega \;+\; \beta m\,\sigma\,e^{-\omega/\sigma} \;-\; \alpha \ln \omega,
    \label{eq:V_eff}
\end{equation}
where the first two terms correspond to the energetic contributions of the abandonment tendency and Gompertz-type early suppression, respectively, while the logarithmic term encodes the entropic growth of accessible configurations with \(\omega\). This potential can also be viewed as the single-particle confining potential in the one-dimensional Coulomb gas representation o the Wishart ensemble. The first two terms act as large-$\omega$ confinment (the linear term being the standard Wishart term, while the exponential piece introduces a softer shoulder), while the logarithmic term corresponds to the hard-edge repulsion at the origin.

Given the potential analogy, for this specific model, the potential always forms a confining well with absent Kramer's escape velocity. In other words, a trend or idea can never grow indefinitely, since the second derivative is always positive, and the first derivative guarantees a root when:
\begin{equation}
    V'(\omega) =0\rightarrow \beta m \left[ 1 - \exp\left(-\omega/\sigma\right) \right] = \frac{\alpha}{\omega}.
\end{equation}

\begin{figure}[H]
    \centering
    \includegraphics[width=0.5\linewidth, angle=270]{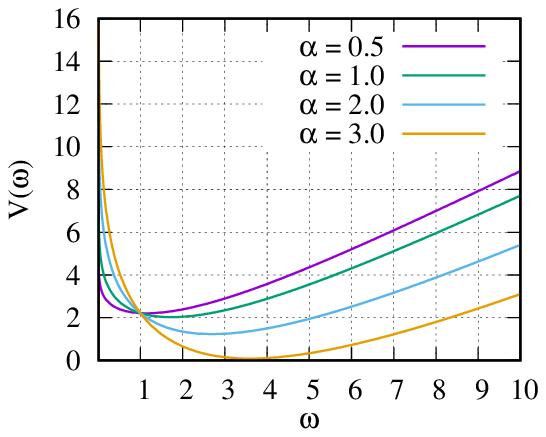}
    \caption{Effective potential 
    \(
    V(\omega) = \beta m\,\omega + \beta m\,\sigma\,e^{-\omega/\sigma} - \alpha \ln\omega
    \)
    for various values of \(\alpha\), with fixed parameters \(\beta=1.0\), \(m=1.0\), and \(\sigma=2.0\).
    Increasing \(\alpha\) shifts the position of the global minimum \(\omega_\ast\) to larger values, corresponding in the innovation model to stronger entropic reinforcement effects in the adoption dynamics.}
    \label{fig:potential_well}
\end{figure}

The well-like potential is shown in Fig. \ref{fig:potential_well} for different values of $\alpha$.

To describe the deterministic relaxation of an innovation within this potential landscape, we adopt a mechanical analogy in which \(\omega\) is the coordinate of a unit-mass particle evolving under \(V(\omega)\).  
The corresponding Lagrangian is:
\begin{equation}
    \mathcal{L}(\omega, \dot{\omega}) = \frac{1}{2} \dot{\omega}^{2} - V(\omega),
    \label{eq:Lagrangian}
\end{equation}
where the kinetic term models the inertial persistence of the innovation state with respect to \(\omega\), and \(V(\omega)\) acts as the generalized adoption–abandonment potential.

Applying the Euler–Lagrange equation:
\[
\frac{d}{dt} \left( \frac{\partial \mathcal{L}}{\partial \dot{\omega}} \right)
- \frac{\partial \mathcal{L}}{\partial \omega} = 0
\]
to Eq.~\eqref{eq:Lagrangian} yields the equation of motion:
\begin{equation}
    \ddot{\omega} = -\beta m \left[\, 1 - e^{-\omega/\sigma} \,\right] + \frac{\alpha}{\omega}.
    \label{eq:EL_eqn}
\end{equation}

The first term on the right-hand side represents the net ``force'' from the energetic component of the potential: for small \(\omega\), it is reduced by the Gompertz barrier; for large \(\omega\), it approaches a constant abandonment pull.  
The second term is a purely entropic force arising from the degeneracy factor, repulsive for \(\alpha > 0\) and attractive for \(\alpha < 0\). 

\begin{figure}[H]
    \centering
    \includegraphics[width=0.8\linewidth]{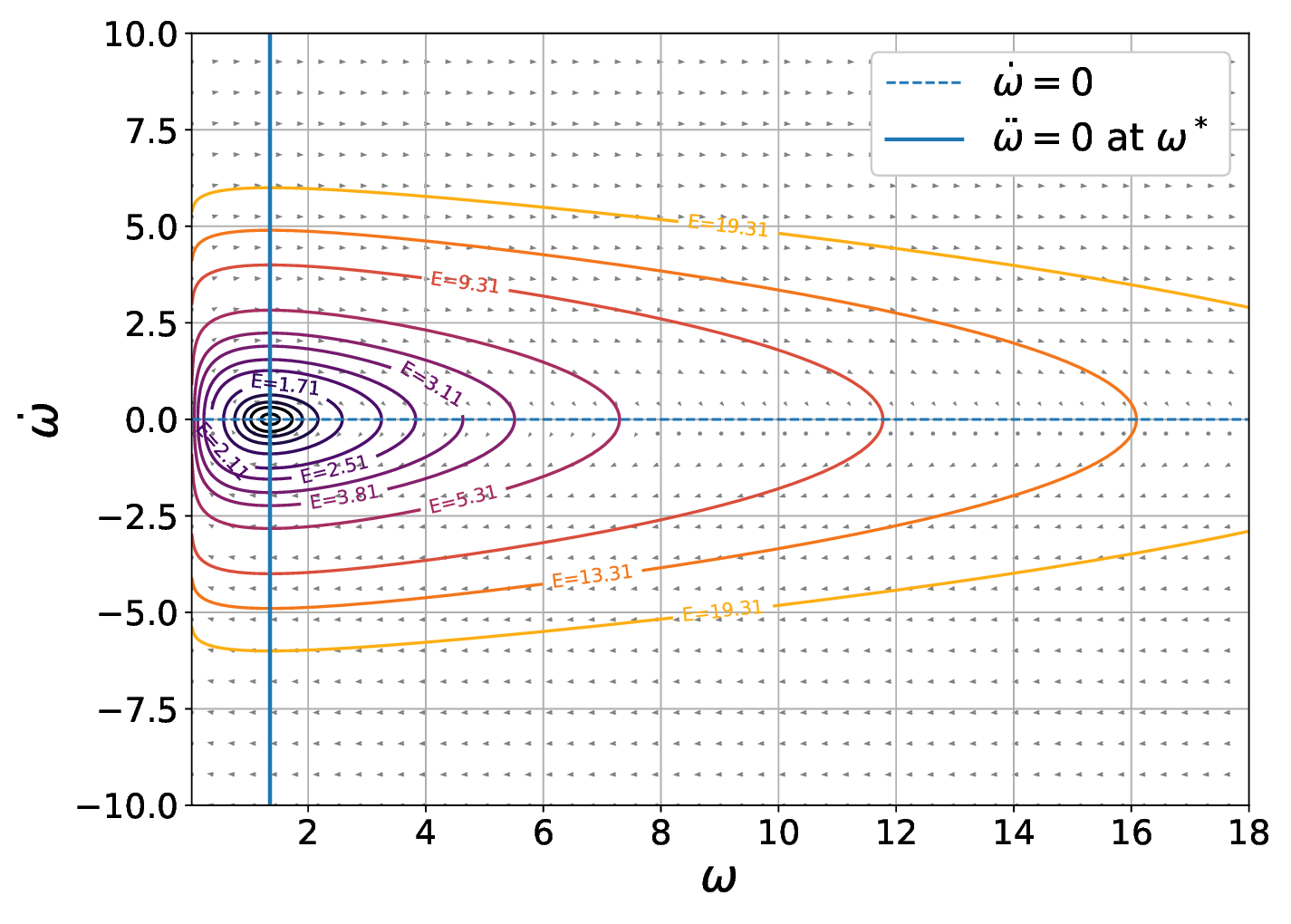}
    \caption{Phase–space representation of the system $\ddot{\omega} = -\beta m \left[ 1 - e^{-\omega/\sigma}\right] + \alpha/\omega$. Contours correspond to constant energy $H(\omega,\dot{\omega})=\frac{1}{2} \dot{\omega}^{2} + U(\omega)$, highlighting both bounded and unbounded trajectories. The equilibrium point $(\omega^{*},0)$ is marked by the intersection of the nullclines $\omega^{*}=0$ and $\ddot{\omega}=0$, while the quiver field shows the direction of phase–space flow.}
    \label{fig:phase-space}
\end{figure}

The corresponding phase–space portrait, shown in Fig. \ref{fig:phase-space}, illustrates the trajectories in the $(\dot{\omega},\omega)$ plane generated by these competing forces. Energy level contours reveal the closed orbits near the equilibrium point and the open trajectories at higher energies, while the vector field indicates the local flow direction.

Equation~\eqref{eq:EL_eqn} thus describes the effective adoption–abandonment dynamics in the generalized coordinate \(\omega\), where the interplay of early-stage suppression, large-\(\omega\) decay, and entropic effects shapes the system's trajectory in analogy with a mechanical particle moving in a non-linear potential. This framework is agnostic to the physical meaning of \(\omega\), making it applicable to time-driven, exposure-driven, or other progress-based formulations.

The well-like potential in Fig. \ref{fig:potential_well} and the mechanistic analogy in Fig. \ref{fig:phase-space} show a potential with a single global minimum, which means that there is no phase-transition. Investigating further, however, we see that Eq. \ref{eq:Innovation_Dynamics} contains two important terms: a degeneracy factor $\omega^{\alpha}$ that shifts the curve towards larger $\omega$ values and an exponential cutoff term that may or may not have a suppression effect on initial adoption. Both of these terms counterbalance each other mediated by innovation temperature $\beta$. Thus, depending on the magnitude of $\beta$, two distinct thermodynamic profiles emerge even without a real potential barrier being introduced in the system. This phenomena will be further discussed in the next session with real-world data to corroborate this conclusion.

\section{Thermodynamic Interpretation}

Within the canonical mapping, the model parameters admit a direct thermodynamic interpretation that is independent of the physical meaning assigned to $\omega$.  
The exponent $\alpha$ specifies the entropic degeneracy factor, determining how the number of accessible microstates scales with the progress coordinate $\omega$.  
The parameter $\beta$ plays the role of an inverse temperature, setting the relative weighting of low- versus high-energy states in the ensemble.  
The characteristic scale $\sigma$ controls the crossover between early-stage suppression and asymptotic abandonment, while $m$ fixes the global energy scale of both contributions to $E(\omega)$.

Thermodynamic observables follow from derivatives of the logarithm of the partition function~\eqref{eq:partition_function}. These observables are essential to connect the real-world datasets with the model and determine the thermodynamical analogies of these systems. Starting with ensemble-averaged energy:
\begin{equation}
\langle E \rangle = -\frac{\partial}{\partial \beta} \ln Z(\alpha,\beta,m,\sigma),
\label{eq:mean_energy}
\end{equation}
which in this framework represents the average energetic cost associated with maintaining the innovation state at a given $\omega$.  

Heat-capacity is one of the most important response functions to study a thermodynamic system. It provides knowledge on the sensitivity of energy due to fluctuations in temperature, and given the analogy to innovation systems, a large heat capacity indicates that small changes in exposure lead to large fluctuations in adoption intensity, i.e. the system is highly adaptable but also volatile. Conversely, a small heat capacity signals rigidity and insensitivity to change. We define it as: 
\[
    C = \frac{\text{Var}(E)}{k_{B}T^{2}} \, ,
\]
since our framework does not include the Boltzmann constant, we adopt units where $k_{B}=1$. Accordingly, we instead interpret the temperature simply as $T = 1/\beta'$ which simplifies the heat capacity to:
\begin{equation}
    C = \text{Var}(E) \beta'^{2} \, .
\end{equation}

Another frequently utilized response function, is the susceptibility, which is a quantification of how sensitive the system is under changes in an externally applied field. To establish a relationship of the model with a magnetic system under the influence of an external field, we write its partition function:
\[
    Z(h) = \sum_{\{s\}} \exp\left(\beta h M(\{s\}) - \beta H_{0}(\{s\})\right) \, ,
\]
where $M(\{s\})$ is the magnetization of the microscopic state $\{s\}$, $h$ is the external field, $\beta$ is the inverse of the magnetic system's temperature, and $H_{0}$ is the system's bare Hamiltonian when no external field is applied. The analogy in our innovation model follows the same structure, with the partition function taking form of Eq. \ref{eq:partition_function}, which allows the identification of $\alpha$ as the external field, and $\ln(\omega)$ as magnetization, as well as $\left(\sigma \exp(-\beta/\sigma)+\omega\right)$ taking the form of the bare Hamiltonian. Given the relationships, the susceptibility becomes:
\[
    \chi_{\alpha} = \frac{\partial M}{\partial h} = \frac{\partial \langle \ln(\omega) \rangle}{\partial \alpha} = \text{Var}\big(\ln(\omega)\big) \, .
\]

In the innovation analogy, susceptibility $\chi_{\alpha}$ measures relative diversity in adoption, in other words, a large $\chi_{\alpha}$ indicates that small changes in the ``diversity pressure'' $\alpha$ lead to disproportionately large fluctuations in adoption distribution, i.e., the system is in a volatile regime where adoption patterns are highly uneven, so some exposure levels $\omega$ can lead to very large adoption levels while others may hardly trigger adoption.

While heat capacity and susceptibility capture the magnitude of fluctuations in adoption behavior, they do not reveal whether adoption responses are biased toward early or late adoption. To capture this asymmetry, we also computed the Bowley skewness of the adoption distribution \cite{Bowley1901}. Bowley's skewness is the difference between a distribution's quartiles:
\begin{equation}
    \tilde{\mu}_{3} =  \frac{Q_{3} + Q_{1} - 2Q_{2}}{Q_{3}-Q_{1}} \, ,
\end{equation}
where each $Q_{i}$ represents the $i^{\text{th}}$ quartile of an arbitrary curve. Negative skewness $\tilde{\mu}_{3}$, indicates that small exposures disproportionately favor early adoption, whereas positive skewness reflects systems where most adoption occurs only at higher exposures. This measure thus complements our thermodynamic observables by characterizing the directional bias of adoption volatility. A necessary measure to establish the type of adoption regime the system finds itself at: slow but progressive Socractic adoption or fast and disruptive Schumpeterian adoption.

While the previous quantities highlight fluctuations and directional biases, we also derive entropy, which provides a complementary global measure, quantifying the overall uncertainty and diversity of adoption pathways within the system. It can describe how widely adoption is distributed across possible states, serving as an indicator of systemic unpredictability, we define it as:
\begin{equation}
S = -\int_{0}^{\infty} P(\omega)\,\ln P(\omega)\,d\omega,
\quad
P(\omega) = \frac{\omega^{\alpha} e^{-\beta m (\omega + \sigma e^{-\omega/\sigma})}}{Z}.
\label{eq:entropy}
\end{equation}
It measures the logarithmic spread of accessible adoption–abandonment configurations across the space of $\omega$.  

For large $\sigma$, the entropy approaches that of a Maxwell–Boltzmann ensemble with density-of-states factor $\omega^{\alpha}$, meanwhile, finite $\sigma$ reduces the entropy due to the presence of the Gompertz-type suppression term. The full probability distribution $P(\omega)$ encodes three distinct adoption regimes predicted by the model:
\begin{enumerate}
    \item \textbf{Early suppression} for $\omega \ll \sigma$, dominated by the exponential barrier;
    \item \textbf{Mid-scale acceleration} where entropic growth of available configurations overcomes abandonment forces;
    \item \textbf{Terminal decay} at $\omega \gg \sigma$, governed by the linear energy term.
\end{enumerate}
This is also clear from the potential $V(\omega) \times \omega$ plots shown in Fig. \ref{fig:potential_well} which form a well for some specific $\omega^*$, marking the mid-scale acceleration, while it forms an early barrier for small $\omega$ values and another barrier for large $\omega$. Meanwhile the effects on the biasedness towards early or late adoption are controlled by a change in innovation temperature $T=1/\beta$.

To validate the thermodynamical connection with innovation models established by our model, we fit real datasets to Eq. \ref{eq:Innovation_Dynamics} and determine the parameters $M_{0},\alpha, \beta, \sigma$ using the usual Levenberg-Marquardt fitting technique \cite{daCunha2024}.  

In the context of real datasets, we interpret the generalized variable $\omega$ as exposure, defined as the cumulative sum of adoption events up to time step $t$:
\[
    \omega(t) = \sum_{\tau=0}^{t} M(\tau),
\]
where $M(\tau)$ is the measured adoption at discrete time $\tau$. While in the theoretical formulation $\omega$ may be considered an abstract progress coordinate, for empirical validation we identify it with accumulated exposure, i.e., the total adoption up to that point. This discrete formulation directly matches the structure of the available datasets and ensures that the fitted parameters capture the observed diffusion dynamics.

We apply this procedure both to datasets previously analyzed in \cite{Giardini2024} and to nine additional sources of data. In contrast to earlier work, we now emphasize the thermodynamic interpretation, which strengthens the model’s utility for adoption and abandonment analysis as well as anomaly detection. The datasets include adoption curves of social media platforms such as Facebook, Snapchat, and MySpace, covering a period of 20 years \cite{Giardini2024}, as well as citation curves for 18 physics articles spanning 44 years (from 1980 to 2024) \cite{Altshuler1980,Ambegaokar1980,Brecher1980,Cheng1980,Dasgupta1980,Hashin1980,Keller1980,Klein1980,Koren1980,Kraut1980,Langreth1980,Leguillou1980,Magg1980,Maradudin1980,Mizushima1980,Shifman1980,Su1980,Wiscombe1980}. The resulting fits demonstrate that, even in real-world scenarios, the statistical–mechanical analogy of innovation is robust. Moreover, we observed that the systems can be clustered into four distinct innovation regimes, arising from the combination of two complementary classifications. In terms of rise and decay shape, Maxwell–Boltzmann–type curves ($\omega/\sigma \ll 1$) show smooth, diffusion-like adoption, while Gompertzian curves ($\omega/\sigma \sim 1$) present sharp inflection and asymmetric decay. In parallel, the skewness of the fitted distributions distinguishes Schumpeterian profiles, characterized by rapid, concentrated bursts of adoption, from Socratic profiles, in which adoption spreads more slowly and heterogeneously. Taken together, these two dimensions yield four possible classes of adoption dynamics: Maxwell–Boltzmann–Schumpeterian, Maxwell–Boltzmann–Socratic, Gompertzian–Schumpeterian, and Gompertzian–Socratic.

In \cite{Giardini2024}, most of the utilized curves were of the Schumpeterian types, with technologies and ideas that were disruptive in a short period of time as it is expected of popular social media platforms and well cited publications, however, the other aforementioned process of adoption is also very important, sometimes individuals may not be ``ready'' to adopt some idea, or technology may not be advanced enough to prove some proposed research theory, so initial adoption can be slow. Since the two kinds of adoption profiles show different thermodynamic characteristics, we utilize a larger dataset than in the previous study, including more adoption curves consisting on Socratic type adoption.

\begin{figure}[h!]
    \centering    
    \includegraphics[width=0.6\linewidth, angle=270]{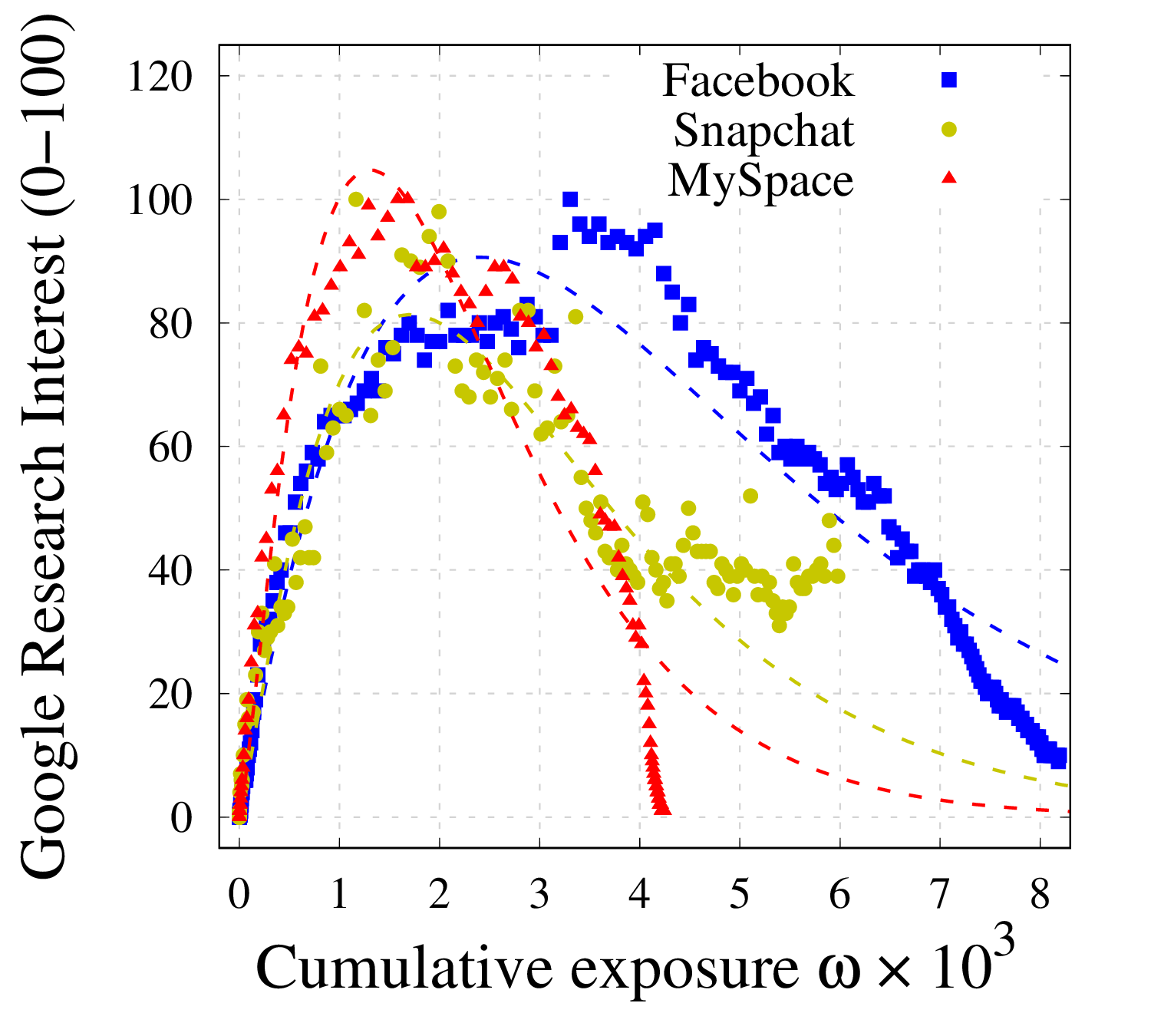}
    \caption{Adoption curves extracted from Google Trends \protect\footnotemark, representing normalized research interest (0–100) for MySpace, Snapchat, and Facebook. The dashed lines show the empirical datasets, while the dashed lines correspond to the theoretical fits obtained using Eq. \ref{eq:Innovation_Dynamics}. }
    \label{fig:social_media_adoption}
\end{figure}

\footnotetext{\url{https://trends.google.com/trends/}} 

Figure \ref{fig:social_media_adoption} shows the adoption curves and corresponding fits for social media, with their thermodynamic quantities described in Tab. \ref{tab:Social_Media_thermodynamic_variables_table}. As expected, the internet and fast communication among individuals, lead to a rapid initial adoption in all of the social media curves, corresponding to a Schumpeterian adoption profile and characterized by a positive skewness. All three curves show periods where the models start to disagree with the adoption curves. Our understanding, also highlighted in \cite{Giardini2024} is that the disagreement originates from anomalies. Both Snapchat and Facebook curves exhibit a long tail divergence that agrees with the onset of the COVID-19 epidemic, where policies of isolation were implemented and the use of social media unequivocally changed for most of the world population \cite{Cinelli2020}. Both curves show a divergence during a period very close to November 2019, when China first reported the arising of a new and very infectious disease \cite{Spiteri2020}. The second anomaly occurs in the MySpace adoption curve, right after reaching peak adoption, Facebook surpassed it in Google searches for the first time in 2008 ($50\sim 60$ months after its creation in 2003), nearly five years after the inception of MySpace, a period where abandonment becomes much stronger and in a market where winner-takes-all competition occurs \cite{Becker1999,Ribeiro2014}. In other words, the second anomaly is not caused by simple abandonment due to individuals natural disinterest curves, but due to competition with another technology that essentially replaced it. 

\begin{table}[h!]
\centering
\scriptsize
\setlength{\tabcolsep}{4pt}
\renewcommand{\arraystretch}{1.2}
\begin{tabular}{r|c c c c c c c c c c c c}
Platform & $M_{0}$ & $\alpha$ & $\beta$ & $\sigma \; \times 10^{-3}$ & $T$ & $Z$ & $E_{\text{mean}}$ & $S$ & $C$ & $\chi_{\alpha}$ & Skew & RMS of Res. \\
\hline
MySpace  & 0.0257 & 1.3475 & 0.2551 & 4.06 & 3.9203 & 29.67  & 9.20 & 3.05 & 2.35 & 0.529 & 0.157 & 13.40 \\
Snapchat & 0.0321 & 1.2172 & 0.6636 & 1.08 & 1.5069 & 2.76   & 3.34 & 2.06 & 2.22 & 0.567 & 0.162 & 8.41  \\
Facebook & 0.0559 & 1.0893 & 2.8932 & 0.16 & 0.3456 & 0.113  & 0.72 & 0.54 & 2.09 & 0.605 & 0.167 & 7.95  \\
\hline
\end{tabular}
\caption{Fitted parameters and derived thermodynamic quantities for the adoption curves of major social-media platforms, obtained from Google Trends data\protect\footnotemark. The adoption dynamics were fitted using Gnuplot’s standard Levenberg–Marquardt algorithm, following Eq. \ref{eq:Innovation_Dynamics}. In addition to the fitted parameters ($M_{0}$, $\alpha$, $\beta$, $\sigma$), the table reports the corresponding thermodynamic quantities—temperature ($T$), partition function ($Z$), mean energy ($E_{\text{mean}}$), entropy ($S$), heat capacity ($C$), susceptibility with respect to $\alpha$ ($\chi_{\alpha}$), and Bowley skewness. Each quantity was calculated using the formalism derived in this work. The RMS of residual quantifies the quality of the fit.}
\label{tab:Social_Media_thermodynamic_variables_table}
\end{table}

\footnotetext{\url{https://trends.google.com/trends/}}

While the social media datasets provide a clear testbed for our framework, their trajectories are strongly affected by anomalies such as the global disruption of COVID-19 and winner–takes–all competition. Such features are intrinsic to platform adoption but may raise the concern that the observed thermodynamic signatures simply do not reflect certain dynamics of adoption. To address this, we turn to a second adoption system: scientific citations. In contrast to social media, where dominance by a single platform can abruptly reshape the landscape, citation growth is generally cumulative and additive, with individual contributions building upon one another. Only in rare cases (two competing theories clash and one is validated) does a winner–takes–all scenario emerge. Thus, we may classify this second scenario as a pure innovation case, without anomalies as constant as in the social-media scenario.

\begin{figure}
    \centering
    \includegraphics[width=0.6\linewidth, angle=270]{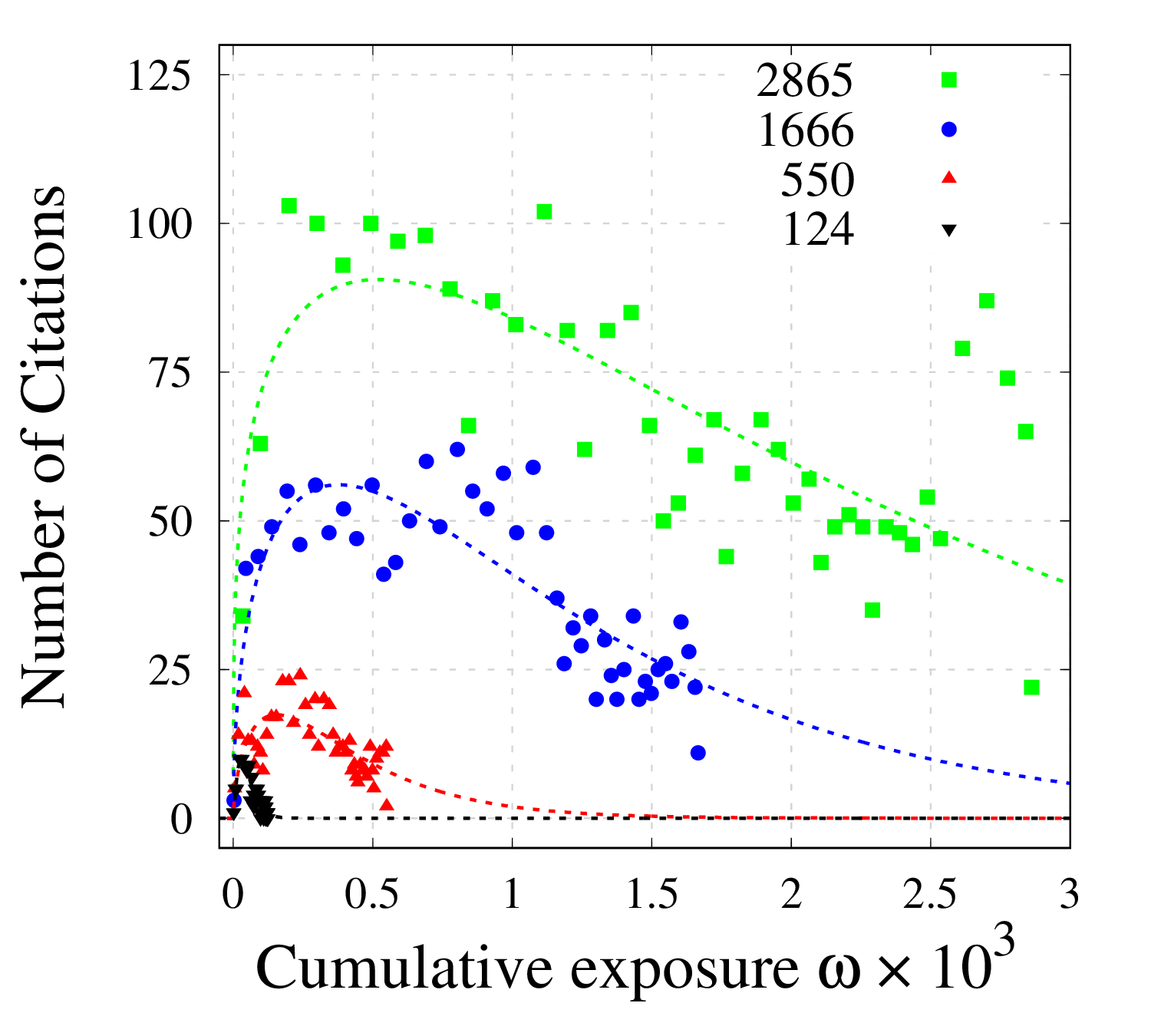}
    \caption{Adoption curves of four representative papers exhibiting Schumpeterian-type growth, with citation metrics extracted from Scopus\protect\footnotemark. Each dataset is fitted with the innovation model given in Eq. \ref{eq:Innovation_Dynamics}. The four examples were selected from the nine available Schumpeterian-type datasets to provide a representative yet uncluttered illustration of the overall adoption dynamics.}
    \label{fig:gompertzian_citations}
\end{figure}

\begin{figure}[h!]
    \centering
    \includegraphics[width=0.6\linewidth, angle=270]{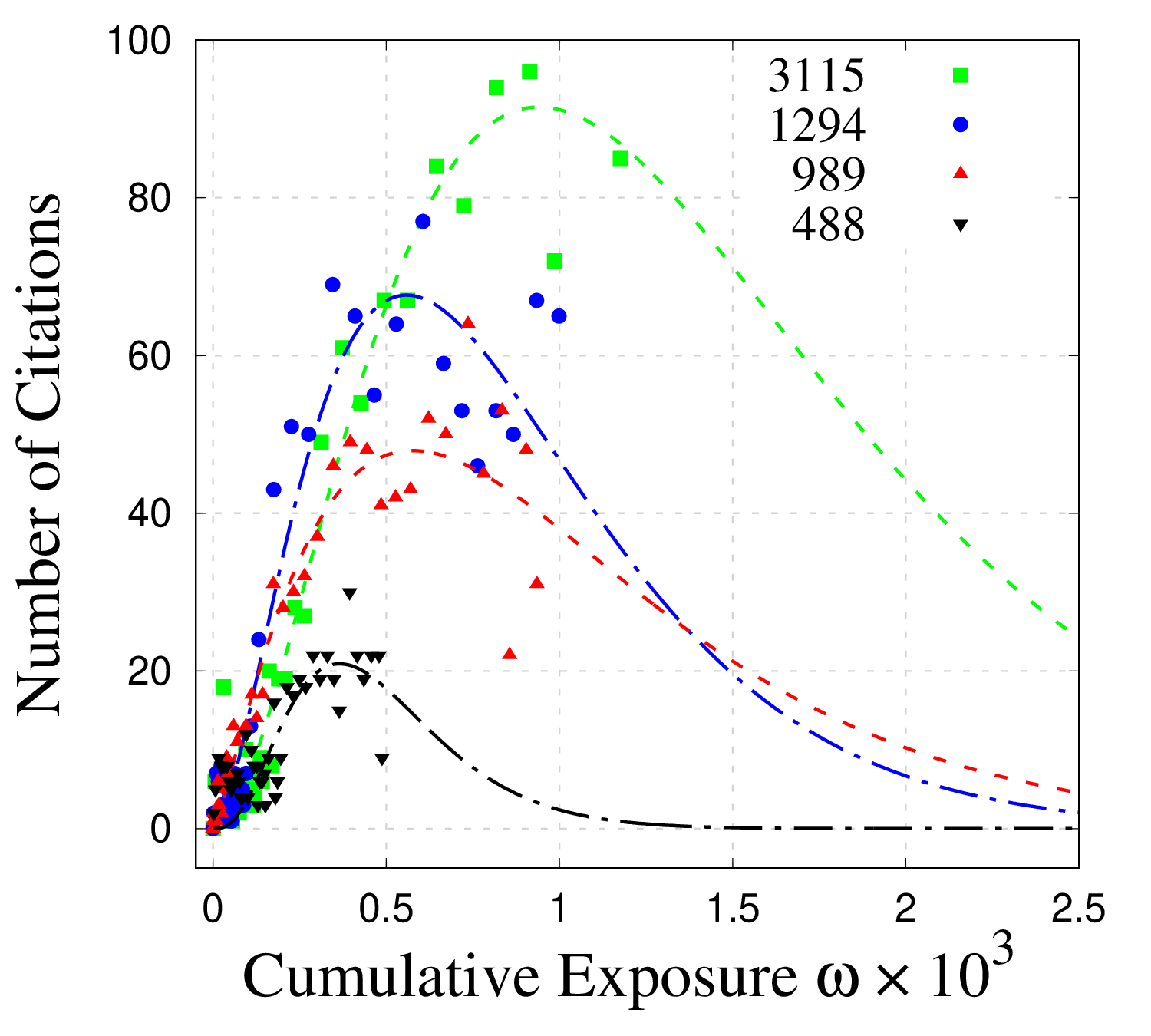}
    \caption{Adoption curves of four representative papers exhibiting Socratic-type growth, with citation metrics extracted from Scopus\protect\footnotemark. Each dataset is fitted with the innovation model of Eq. \ref{eq:Innovation_Dynamics}. The four examples were selected from the nine available Socratic-type datasets to provide a comprehensive yet uncluttered illustration of the overall adoption dynamics.}
    \label{fig:Maxwell-Boltzmann_citations}
\end{figure}

We therefore extend the analysis to citation profiles, which offer a complementary setting largely free from the disruptive anomalies seen in social media. By applying the same fitting procedure to citation data, we can evaluate the model under conditions where adoption is primarily cumulative and additive, providing a clearer view of the underlying innovation dynamics. Figs. \ref{fig:gompertzian_citations} and \ref{fig:Maxwell-Boltzmann_citations} show the two distinct thermodynamic profiles characterized by rate of initial adoption (Schumpeterian vs Socratic), quantified by skewness of the curve, and defined by two distinct thermodynamic profiles.

\footnotetext{\url{https://www.scopus.com/pages/home?display=basic#basic}}

\begin{figure}[h!]
    \centering
    \includegraphics[width=0.6\linewidth, angle=270]{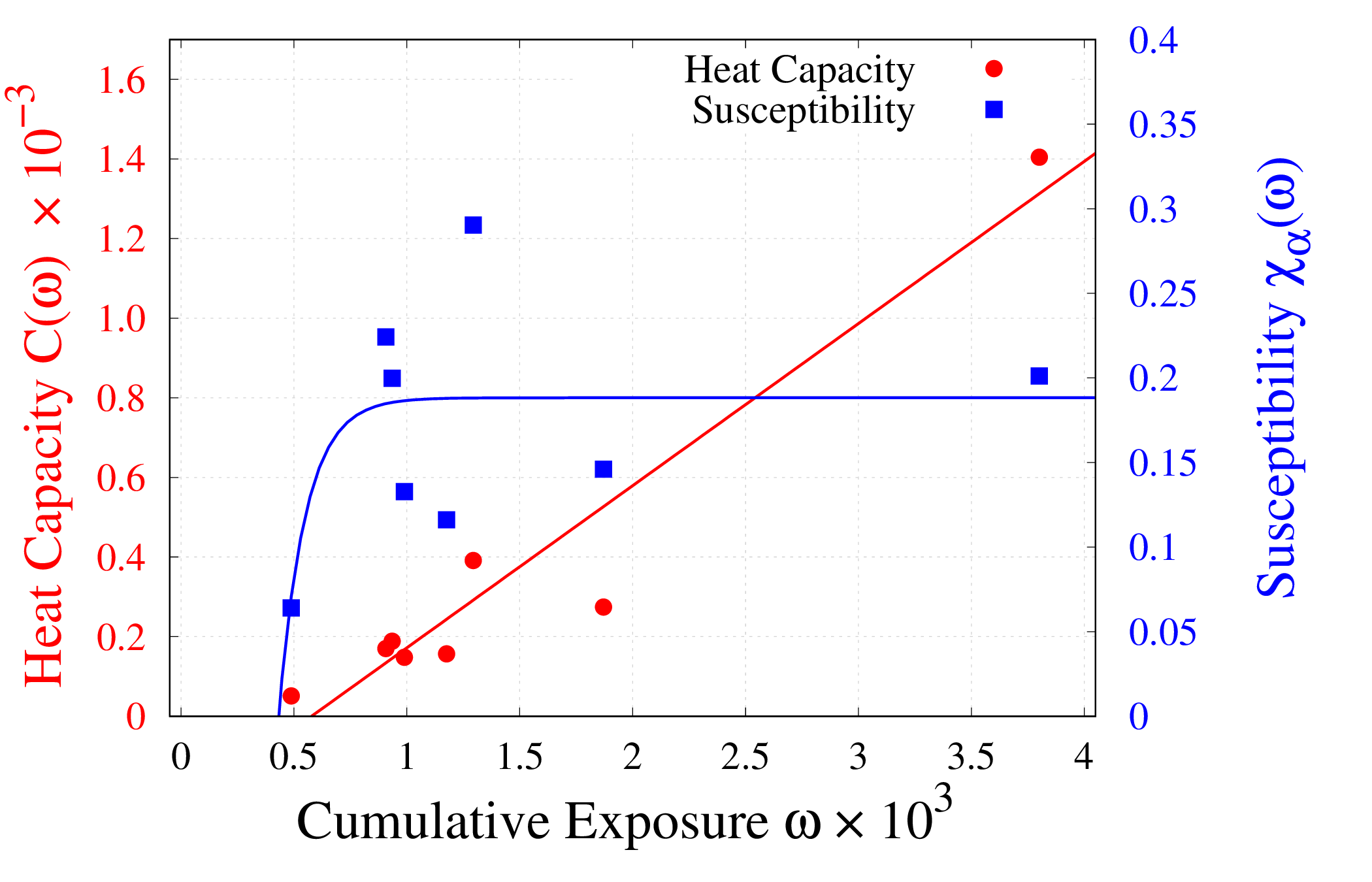}
    \caption{Heat capacity (red, left axis) and susceptibility (blue, right axis) as functions of the adoption level $\omega$ for datasets characterized by Socratic-type growth. The numerical values associated with these curves are provided in Tab. \ref{tab:MB_thermodynamic_variables_table}.}
    \label{fig:MB_citations_response}
\end{figure}

\begin{figure}[h!]
    \centering
    \includegraphics[width=0.6\linewidth, angle=270]{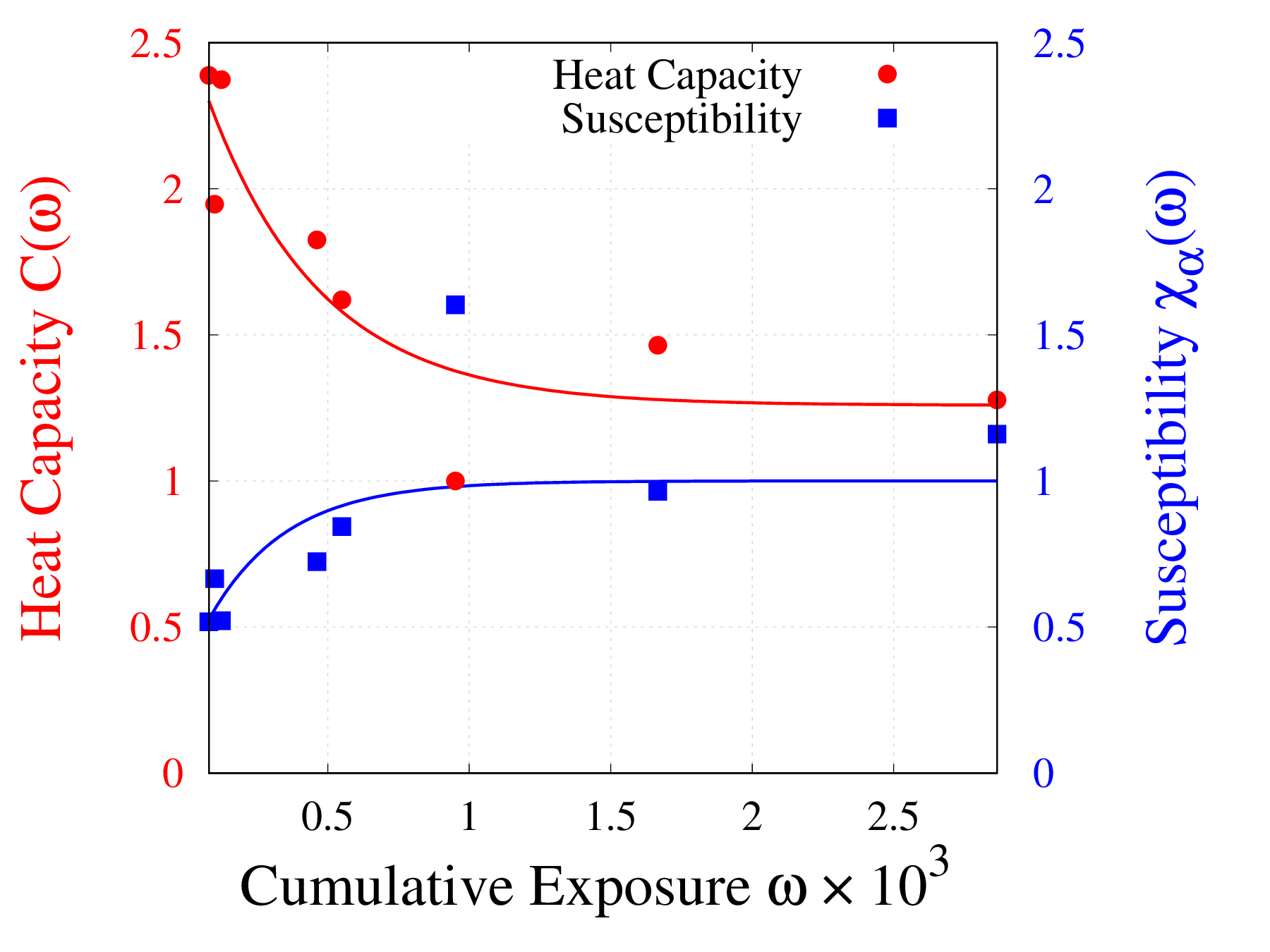}
    \caption{Heat capacity (red, left axis) and susceptibility (blue, right axis) as functions of the adoption level $\omega$ for datasets exhibiting Schumpeterian-type growth (fast initial adoption) with their results displayed in Tab. \ref{tab:GP_thermodynamic_variables_table}.}
    \label{fig:GP_citations_response}
\end{figure}

The Schumpeterian dynamics shown in Fig. \ref{fig:gompertzian_citations} exhibit positive skewness, where the majority of the curve's mass is localized in the first quadrants, an important classification in which innovation heat-capacity and total adoption remain anti-correlated. By contrast, the Socratic citation curves in Fig. \ref{fig:Maxwell-Boltzmann_citations} display negative skewness, where thermal capacity is often positively correlated to total adoption as shown in Fig. \ref{fig:MB_citations_response}. In both cases, the average ensemble energy ( Eq. \ref{eq:mean_energy}) and innovation temperature increase monotonically with exposure $\omega$, reflecting the fact that as adoption grows, the system is driven into progressively higher-energy states—energy expenditure being a prerequisite for adoption to occur. Likewise, entropy grows steadily with exposure across all systems, showing that adoption is not only energetically costly but also expands the configurational space of possibilities. The crucial difference lies in the behavior of heat-capacity: Fig. \ref{fig:GP_citations_response} shows that in the Schumpeterian case, rapid adoption produces a very large initial heat-capacity that declines as exposure grows, indicating strong but saturating responsiveness. Figure \ref{fig:MB_citations_response} shows the Socratic case, wherein gradual adoption yields much smaller absolute heat-capacities, which increase with exposure but never approach the absorptive capacity of the fast-growth regime. This distinction highlights that while energy, entropy, and temperature growth are universal properties of adoption, the sign of skewness determines whether the system begins with high but declining resilience (Schumpeterian) or with low but gradually increasing responsiveness (Socratic) as innovation spreads.

In addition to energy, temperature, and heat-capacity, the susceptibility of adoption also provides an important thermodynamic diagnostic. In both social media and citation systems, we observe that susceptibility is not monotonic but instead peaks at intermediate levels of exposure, as already noted for both Schumpeterian and Socratic dynamics. This peak indicates that the system is most sensitive to external stimuli precisely at the tipping point of adoption: when exposure is still limited, additional influence has little effect, while once adoption saturates, further exposure yields diminishing returns. The intermediate maximum therefore marks the regime where innovation is most volatile to shocks, whether these are competitive pressures in social media or bursts of attention in scientific citations. However, the magnitude of this peak differs strongly across skewness classes: in the fast-growing Schumpeterian case, susceptibility is significantly larger, as seen in the comparison between Tables \ref{tab:MB_thermodynamic_variables_table} and \ref{tab:GP_thermodynamic_variables_table}, highlighting the much greater responsiveness of rapid-adoption systems to external perturbations. In contrast, the Socratic case exhibits a similar mid-exposure peak but at a smaller scale, reflecting its comparatively muted sensitivity throughout the diffusion process.

\begin{table}[h!]
\centering
\scriptsize
\setlength{\tabcolsep}{4pt}
\renewcommand{\arraystretch}{1.2}
\begin{tabular}{r|c c c c c c c c c c c c}
\hline
Cit. & $M_{0}$ & $\alpha$ & $\beta\; \times 10^{-5}$ & $\sigma$ & $T$ & $Z$ & $E_{\text{mean}}$ & $S$ & $C \; \times 10^{-4}$ & $\chi_{\alpha}$ & Skew & RMS of Res. \\
\hline
488  & $9.88\times 10^{-6}$ & 2.9721 & $9.18$ & 88.64 & 10895.77 & $1.16\times 10^{10}$ & 391.16  & 5.56 & $0.511$ & 0.0641 & -0.196 & 4.96  \\
907  & 0.1581               & 1.1294 & $6.24$ & 34.76 & 16016.70 & $8.98\times 10^{5}$  & 614.53  & 6.59 & $1.70$  & 0.224  & -0.137 & 6.68  \\
935  & 0.0570               & 1.2572 & $6.54$ & 33.34 & 15279.67 & $2.16\times 10^{6}$  & 645.12  & 6.59 & $1.89$  & 0.200  & -0.144 & 5.73  \\
989  & 0.00552              & 1.7683 & $6.00$ & 52.79 & 16656.03 & $6.95\times 10^{7}$  & 731.50  & 6.53 & $1.48$  & 0.133  & -0.166 & 7.53  \\
1176 & 0.000956             & 1.9614 & $5.43$ & 38.38 & 18418.18 & $3.99\times 10^{8}$  & 876.27  & 6.65 & $1.57$  & 0.116  & -0.172 & 6.61  \\
1294 & 0.2300               & 0.8783 & $6.29$ & 8.97  & 15897.11 & $3.54\times 10^{5}$  & 838.24  & 7.01 & $3.91$  & 0.290  & -0.118 & 7.41  \\
1871 & 0.00169              & 1.6472 & $4.25$ & 11.45 & 23527.87 & $1.64\times 10^{8}$  & 1351.58 & 7.19 & $2.74$  & 0.146  & -0.161 & 10.57 \\
3801 & 0.0267               & 1.2844 & $4.38$ & 6.84  & 22821.17 & $5.88\times 10^{7}$  & 2612.04 & 7.99 & $14.04$ & 0.201  & -0.139 & 5.97  \\
\hline
\end{tabular}
\caption{Thermodynamic characterization of citation datasets following Socratic-type adoption–abandonment dynamics. Citation counts were obtained from Scopus and fitted with the innovation model of Eq.~\ref{eq:Innovation_Dynamics} using the Levenberg–Marquardt algorithm. Beyond the primary fit parameters ($M_{0}$, $\alpha$, $\beta$, $\sigma$), the table highlights emergent thermodynamic indicators—temperature ($T$), partition function ($Z$), mean energy ($E_{\text{mean}}$), entropy ($S$), heat capacity ($C$), susceptibility ($\chi_{\alpha}$), and skewness—that quantify stability, diversity, and responsiveness of the adoption trajectories. The RMS of residuals provide a measure of goodness-of-fit.}
\label{tab:MB_thermodynamic_variables_table}
\end{table}

\begin{table}[h!]
\centering
\scriptsize
\setlength{\tabcolsep}{4pt}
\renewcommand{\arraystretch}{1.2}
\begin{tabular}{r|c c c c c c c c c c c c}
\hline
Cit. & $M_{0}$  & $\alpha$             & $\beta$ & $\sigma$ & $T$   & $Z$                 & $E_{\text{mean}}$ & $S$   & $C$    & $\chi_{\alpha}$ & Skew & RMS of Res. \\
\hline
80   & 0.09661  & 1.3885               & 0.7591  & 0.05156  & 1.317 & 2.381               & 3.147             & 1.975 & $2.39$ & 0.518           & 0.155    & 1.32  \\
100  & 0.3602   & 0.9474               & 0.6208  & 0.04788  & 1.611 & 2.477               & 3.137             & 2.035 & $1.95$ & 0.666           & 0.174    & 1.67  \\
124  & 0.3772   & 1.3738               & 0.3847  & 0.1214   & 2.599 & 11.79               & 6.170             & 2.650 & $2.37$ & 0.522           & 0.156    & 1.28  \\
462  & 0.3609   & 0.8253               & 0.1401  & 0.02134  & 7.138 & 33.92               & 13.03             & 3.477 & $1.83$ & 0.724           & 0.181    & 4.66  \\
550  & 1.411    & 0.6197               & 0.1219  & 0.03248  & 8.199 & 27.05               & 13.28             & 3.525 & $1.62$ & 0.844           & 0.194    & 4.16  \\
951  & 40.60    & $5.52\times 10^{-6}$ & 0.0744  & 0.01628  & 13.44 & 13.43               & 13.45             & 3.598 & $1.00$ & 1.603           & 0.262    & 8.34  \\
1666 & 5.661    & 0.4646               & 0.0352  & 0.03488  & 28.39 & $1.19\times 10^{2}$ & 41.58             & 4.688 & $1.46$ & 0.965           & 0.206    & 8.37  \\
2865 & 21.07    & 0.2775               & 0.0231  & 0.02305  & 43.30 & $1.11\times 10^{2}$ & 55.31             & 4.996 & $1.28$ & 1.160           & 0.224    & 15.50 \\
\hline
\end{tabular}
\caption{Fitted parameters and derived thermodynamic quantities for citation datasets exhibiting Schumpeterian-type growth, with citation metrics extracted from Scopus. The adoption dynamics were fitted using Gnuplot’s standard Levenberg–Marquardt algorithm, following Eq.~\ref{eq:Innovation_Dynamics}. Reported values include the fitted parameters ($M_{0}$, $\alpha$, $\beta$, $\sigma$) and the corresponding thermodynamic quantities—temperature ($T$), partition function ($Z$), mean energy ($E_{\text{mean}}$), entropy ($S$), heat capacity ($C$), susceptibility with respect to $\alpha$ ($\chi_{\alpha}$), and Bowley skewness. The RMS of residual provides a measure of fit quality across the datasets.}
\label{tab:GP_thermodynamic_variables_table}
\end{table}

Taken together, the results demonstrate that innovation dynamics, when cast into a thermodynamic framework, are both constrained and richly structured. The effective potential derived from the model always has a single minimum, ensuring that the system never undergoes a classical phase transition: there is no Kramer's escape velocity of adoption, and no trend can persist indefinitely without continuous expenditure of innovation energy. 

We show that dynamics are governed by how temperature and degeneracy interact with the potential to shape ensemble behavior. Two distinct thermodynamic patterns emerge: In the first, corresponding to rapid initial adoption with a narrow distribution (Schumpeterian profiles), increasing adoption suppresses fluctuations, reflected in a decrease of the heat-capacity with total exposure. However, the absolute values of heat-capacity in this regime are an order of magnitude larger than in slow-growth cases, meaning that fast adoption—though it rigidifies over time—begins with a very high absorptive capacity that enables it to buffer large shocks during the critical growth phase. In the second, corresponding to slower and broader adoption (Socratic profiles), rising adoption expands the distribution, amplifies fluctuations, and drives the heat-capacity upward. Yet the absolute levels remain much smaller, indicating that this mode of growth never develops the same systemic capacity as its fast-growth counterpart, even if its fluctuations appear steadier.

Beyond this skewness-based classification, the second axis of variability—Maxwell–Boltzmann versus Gompertzian forms—reflects the role of the parameter $\sigma$. Thermodynamically, $\sigma$ controls the effective scale of exposure at which adoption saturates, analogous to a characteristic energy spread in statistical mechanics. For large $\sigma$, the Gompertz term reduces to its Taylor expansion, yielding a Maxwell–Boltzmann–like curve that mirrors the smooth energy distribution of an ideal gas. For small $\sigma$, by contrast, the Gompertz term dominates, introducing sharp inflection points and critical-like behavior, consistent with collective social effects and discontinuous changes in adoption. Given these two regimes, there is significant indication that some innovation systems behave simply like ideal gases, while others become critical and start behaving as Gompertz systems.

Taken together, these results establish the fundamental thermodynamic patterns of adoption and abandonment: single-minimum potentials without phase transitions, skewness-driven differences in heat-capacity, and the role of $\sigma$ in separating Maxwell–Boltzmann–like and Gompertzian regimes. In the next section, we turn from the quantitative characterization of these signatures to their broader interpretation, asking what they imply about the resilience, fragility, and strategic variability of innovation systems.

\section{Discussion and Outlook}

The equilibrium scaffold developed here places innovation–abandonment dynamics on the same footing as statistical mechanics: a generalized state coordinate $\omega$ evolves under an effective potential $V(\omega)$, with associated thermodynamic quantities derived from the partition function $Z$. This construction elevates loose metaphors such as ``momentum'' or ``network fatigue'' into measurable objects. Stability becomes the curvature $\ddot{V}(\omega_\ast)$ at the most probable state, susceptibility corresponds to derivatives of $\ln Z$, and heat-capacity measures volatility of adoption under changing exposure. In this way, innovation dynamics can be compared directly across datasets and domains with a quantitative common language.

One central insight is that the potential $V(\omega)$ possesses a single minimum. As a consequence, innovation trajectories in this framework admit no true phase transition, no runaway escape velocity, and no trend that persists indefinitely without expenditure of innovation energy. Growth can accelerate or stall, but never diverge. This helps explain why adoption curves saturate in practice even when initial growth is explosive.

The interaction of degeneracy, potential, and effective temperature produces two distinct adoption regimes (not to be mistaken for thermodynamical phase-changes). In one, rapid early adoption creates a narrow peak in the distribution and a declining heat-capacity with exposure; yet the absolute values remain far larger than in slow-growth cases, giving these systems a much greater ability to absorb adoption shocks during their critical expansion phase. In the other, slower but broader adoption raises heat-capacity as exposure increases, but from a much lower baseline, so that the system never develops the same absorptive capacity. This suggests that the famous dictum of Silicon Valley, to ``move fast and break things'' may have had an unintentional thermodynamic truth: systems that grow explosively begin with enormous absorptive capacity, whereas for a slow-growth company to reach a comparable level of heat-capacity, it must remain in the market for much longer before accumulating sufficient exposure. Faster growth is therefore not inherently fragile, but structurally endowed with greater thermodynamic resilience, while slower growth is steady only in appearance, remaining comparatively inert. Papers, whose strategies are fixed at publication, naturally fall into the latter regime, whereas adaptive platforms can shift strategies dynamically and thereby preserve high thermodynamic capacity across their life-cycle.

A second distinction emerges from the role of $\sigma$, the novelty-decay scale. Thermodynamically, $\sigma$ sets the spread of adoption states, analogous to an energy scale in gases. For large $\sigma$, the Gompertz correction reduces to a Taylor expansion, yielding Maxwell–Boltzmann–like curves that mirror the smooth spreading of energy in an ideal gas. In this regime, social systems behave like gases, where individuals act as particles exchanging influence and information, and diffusion follows patterns directly comparable to the Maxwell–Boltzmann distribution of energies. For small $\sigma$, however, the Gompertz term dominates, producing sharper inflection points and critical-like social behavior driven by the $1/\omega$ growth factor. Adoption dynamics may then enter critical phases characterized by abrupt tipping points or cascades, consistent with instabilities in physical critical phenomena. Importantly, real social systems may transition between these two regimes, shifting from smooth Maxwell–Boltzmann–like diffusion to critical, cascade-like adoption depending on novelty scale and context. This duality opens new theoretical avenues: large-$\sigma$ cases motivate kinetic-theory approaches, with Boltzmann-like transport equations describing how adoption redistributes across sub-populations, while low-$\sigma$ cases invite kinetic closures capable of capturing criticality in social adoption. Both perspectives point toward the construction of discrete particle models, where agents exchange or withhold ``adoption energy'' as microscopic foundations that can reproduce the observed macroscopic dynamics and, ultimately, toward a non-equilibrium statistical mechanics of innovation.

These thermodynamic contrasts also resonate with sectoral patterns. Slower, steadier adoption resembles ``safe'' sectors such as food or energy, where heat-capacity rises with exposure but only accumulates meaningfully over very long timescales. A company like Nestlé\footnote{https://www.nestle.com/about/history/nestle-company-history}, founded in 1867, exemplifies this trajectory: its longevity allows it to reach levels of absorptive capacity comparable to those of much younger, fast-growth firms. By contrast, high-tech firms such as Facebook achieve extremely high heat-capacity within just a few years of explosive adoption, though their capacity then declines as saturation sets in. Crucially, these regimes are not fixed: firms may transition between them, adopting rapid expansion strategies to capture early markets and later shifting toward steady accumulation to sustain resilience over the long run.

Another important consequence of this model, lies in anomaly detection. As seen in social media adoption, competitive shocks, winner-take-all scenarios, or global events disturb the natural equilibrium trajectory. Within this framework, such events appear as deviations in the adoption profiles, providing a principled way to flag when adoption is being shaped by exogenous shocks rather than internal dynamics or simple natural adoption dynamics.

At the same time, limitations must be acknowledged. The present framework is inherently an equilibrium theory: it describes systems whose adoption rules remain fixed. Many real innovations—especially corporate or technological platforms—alter strategies mid-course, creating new adoption cycles that decouple from earlier ones. Such non-stationarity is outside the scope of the current model, but the equilibrium scaffold provides a baseline against which such departures can be measured.

In summary, by framing adoption–abandonment dynamics as an equilibrium statistical mechanics problem, we obtain rigorous thermodynamic diagnostics alongside interpretable sectoral analogies. The next step is to extend these tools into non-equilibrium territory, where strategies shift, multiple innovations compete, and true phase-like transitions may emerge. This progression—from equilibrium scaffolds to kinetic and adaptive theories—offers a roadmap for making innovation dynamics a predictive science rather than a retrospective metaphor.

\bibliographystyle{unsrt}
\bibliography{main}
\newpage


\end{document}